%% file: IEEE-conference-template-062824.tex
\documentclass[conference]{IEEEtran}
\IEEEoverridecommandlockouts

\usepackage{cite}
\usepackage{amsmath,amssymb,amsfonts}
\usepackage{algorithmic}
\usepackage{graphicx}
\usepackage{textcomp}
\usepackage{xcolor}
\usepackage{booktabs}
\usepackage{multirow}
\usepackage{hyperref}
\def\BibTeX{{\rm B\kern-.05em{\sc i\kern-.025em b}\kern-.08em
    T\kern-.1667em\lower.7ex\hbox{E}\kern-.125emX}}

\begin{document}

\title{FdAudio: MeanFlow-Anchored Fr\'echet-Distance Post-Training for One-Step Text-to-Audio Generation}


\author{
  \IEEEauthorblockN{Kuan-Po Huang$^\star$, Bo-Ru Lu$^\dagger$\thanks{$^\dagger$This work is unrelated to the author’s position at Amazon.}, Ho-Lam Chung$^\star$, Shih-Hsin Wang$^\star$,  Hung-yi Lee$^\star$}
  \IEEEauthorblockA{
    \textit{$^\star$National Taiwan University \quad $^\dagger$Amazon} \\
  }
}

\maketitle

\begin{abstract}
While recent few-step sampling text-to-audio generation models like MeanAudio substantially accelerate generation by modeling average velocities, their strict one-step generation quality still lags significantly behind multi-step counterparts.  
We propose FdAudio to bridge this gap. Unlike MeanAudio, which relies solely on regression against target velocity fields, our post-training approach optimizes the final one-step distribution directly across pre-trained embedding spaces via a multi-representation Fréchet-distance (FD) loss. Crucially, to prevent the multi-step degradation that naive post-training with FD-loss causes, we introduce a MeanFlow consistency objective as a structural anchor. 
Results demonstrate that FdAudio establishes state-of-the-art one-step T2A generation quality among few-step systems, yielding an 11.4\% reduction in FD score and a 28.8\% improvement in FAD score relative to the baseline MeanAudio framework. Notably, we solve FD post-training's naive multi-step degradation issue by proposing the MeanFlow anchor, enabling a 25-step sampling path to maintain high-fidelity audio synthesis that matches or surpasses strong multi-step models at a fraction of their computational latency.
\end{abstract}


\section{Introduction}
\label{sec:intro}
\noindent 
Text-to-audio (T2A) aims to automate the process by synthesizing general sound, from ambient scenes to sound events, directly from a natural language prompt, with applications across games, film, and digital content creation. Because inference is performed far more frequently than training in these practical applications, the inference cost becomes a central concern \cite{passoni2025diffused}. However, state-of-the-art systems are mostly built on diffusion \cite{ddpm} and flow-matching \cite{flow-matching} models~\cite{audioldm2, maa2, tango2, tangoflux, impact, audiomntp, ezaudio, resonate}, which produce high-fidelity audio through an iterative sampling process. The model is evaluated sequentially over tens to hundreds of timesteps, at each time step taking a partially denoised latent and predicting a small update that gradually recovers the final clean audio latents. This sequential loop is the primary source of their high inference latency. 

To remove this bottleneck, recent systems distill or shorten the sampling trajectory for few- or one-step inference, including
ConsistencyTTA~\cite{consistencytta}, AudioLCM~\cite{audiolcm},
AudioTurbo~\cite{audioturbo}, SoundCTM~\cite{soundctm}, AudioDEAR~\cite{audiodear}, and
MeanAudio~\cite{meanaudio}. Although dramatically faster, these few-step models still trail their multi-step counterparts in generation quality. This motivates the central question of our work: \emph{can we substantially improve the generation quality of one-step audio generation?}


A recent promising direction in the image generation domain is to apply Fr\'echet-distance (FD) post-training~\cite{fdloss} to finetune a pretrained one-step generator~\cite{pMF, iMF} so that the distribution of its one-step outputs matches real data in the feature spaces of pretrained representation extractors~\cite{inceptv3, convnet-v2, mae-img, dino-v2, siglip2, clip}. This approach is appealing because it is simple and direct: it only requires precomputed real-data statistics, without teacher distillation, adversarial training, or per-sample regression targets.

We adapt this recipe to audio by computing the FD-loss over pretrained audio encoders, including PANNs~\cite{panns}, PaSST~\cite{passt}, BEATs~\cite{beats}, and AudioMAE~\cite{audiomae}. While this substantially improves one-step audio quality, we find that it severely harms multi-step sampling. For example, as shown in the results of Section~\ref{sec:mf-anchor}, a model post-trained with FD-loss achieves a strong one-step FAD of 1.27, but its FAD nearly triples to 3.61 when sampled with 25 steps. This suggests that FD post-training, when applied alone, encourages the model to optimize only the final one-step output. As a result, the model bypasses the iterative flow-matching trajectory and collapses the continuous denoising path into a direct noise-to-data mapping, making multi-step sampling less effective. 
While strong one-step performance is highly desirable for real-time applications, discarding the multi-step trajectory restricts deployment flexibility.

To prevent this collapse, post-training must preserve the model's denoising trajectory rather than only improving its one-step output. Therefore, we introduce the MeanFlow consistency objective~\cite{meanflow}, which mitigates the collapse issue. By regularizing the average velocity over arbitrary sub-intervals $[r,t]$, MeanFlow helps preserve the property that the same model can be sampled with either one or multiple steps. This is precisely the property that naive FD post-training destroys.

Our use of MeanFlow is different from prior work. MeanFlow~\cite{meanflow} and its audio instantiation MeanAudio~\cite{meanaudio} use MeanFlow as the primary training objective for learning fast one-step generators. In contrast, our goal is to leverage the benefits of FD post-training without sacrificing multi-step sampling. We therefore use the FD-loss to improve the one-step output distribution, while repurposing MeanFlow as a lightweight regularizer that preserves the underlying sampling trajectory and prevents the model from collapsing into a purely one-step mapping.

Building on this, we propose \textbf{FdAudio}\footnote{Demo: \href{https://fdoneaudio.github.io/}{https://fdoneaudio.github.io/}}\footnote{Code: \href{https://github.com/nobel861017/FdAudio}{https://github.com/nobel861017/FdAudio}}.
FdAudio combines FD post-training with MeanFlow regularization: the FD-loss improves one-step generation quality, while the MeanFlow term preserves the velocity field needed for high-quality multi-step sampling.

\noindent Our contributions are:
\begin{itemize}
\item We identify a fundamental conflict between distribution-level post-training and trajectory-based generation: optimizing the FD-loss sharpens one-step output but overwrites the velocity field, so the generator can no longer sample across multiple steps.


\item We resolve this conflict with a MeanFlow consistency anchor: a single set of weights that generates high-quality audio in one step and still samples well with many steps without separate models or retraining.

\item At only 120M parameters, FdAudio reaches state-of-the-art one-step T2A generation on AudioCaps among several few-step models, while matching or surpassing other strong multi-step systems.

\end{itemize}


\section{Related Work}
\label{sec:related}


\subsection{Few-step T2A generation}
\noindent While several diffusion- or flow-matching-based T2A models~\cite{audioldm2, maa2, tango2, tangoflux, impact, audiomntp, ezaudio} synthesize high-quality audio, their iterative sampling processes during inference incur high computational costs and substantial latency. 
Formally, given a conditioning text prompt $c$ and an initial noise sample $x_T \sim \mathcal{N}(0, I)$, the standard inference process requires evaluating a parameterized model $F$ over $T$ discrete timesteps. The sampling trajectory produces a sequence of intermediate states $x_T, x_{T-1}, \dots, x_0$, where each update step is defined as:
\begin{equation}
    x_{t-1} = \Phi(x_t, F(x_t, t, c)), \quad t = T, T-1, \dots, 1
\end{equation}
where $\Phi$ denotes the numerical solver or discrete sampling algorithm. The final synthesized audio output is the terminal state $x_0 = x$. Because $T$ typically ranges from tens to hundreds, this sequential evaluation loop acts as a fundamental bottleneck.
This limitation has driven the development of few-step generation techniques. For instance, ConsistencyTTA \cite{consistencytta} and AudioLCM \cite{audiolcm} apply consistency distillation \cite{consistency} to achieve few-step generation, however, the performance of their one-step generation still significantly lags behind state-of-the-art multi-step T2A models. 
More recently, MeanAudio~\cite{meanaudio} instantiated the aforementioned MeanFlow framework for text-to-audio generation, demonstrating that this alternative training objective can yield relatively better one-step audio quality, though it still falls short of its multi-step counterpart. In this work, we aim to close this quality gap, establishing state-of-the-art performance for one-step T2A generation.

\subsection{MeanFlow one-step generation}
\label{sec:meanflow}
\noindent To overcome the high inference latency of traditional iterative sampling, the MeanFlow framework~\cite{meanflow} is designed to directly learn the average velocity of a generation trajectory over an arbitrary time interval. While standard flow-matching models approximate the instantaneous velocity and therefore require extensive numerical integration across many timesteps, MeanFlow structurally bypasses this bottleneck to enable high-fidelity sampling in a single or very few network evaluations.

Mechanically, this framework operates within the latent space of a pretrained autoencoder. Let $x$ denote a clean data latent representation and let $\epsilon \sim \mathcal{N}(0, I)$ represent a noise sample. Standard flow-matching formulates a linear interpolant as $z_t = (1-t)x + t\epsilon$ for time $t \in [0, 1]$, where $z_0 = x$ represents the true data distribution and $z_1 = \epsilon$ represents the pure noise distribution. The instantaneous velocity along this path is defined as $v = \dot{z}_t = \epsilon - x$.

Instead of predicting $v$, the MeanFlow objective targets the \emph{average} velocity, denoted as $u_\theta(z_t, r, t)$, over a specific interval $[r, t]$. This average velocity is formally defined as $u(z_t, r, t) = \frac{1}{t-r}\int_{r}^{t} v(z_\tau, \tau) d\tau$. Differentiating this definition with respect to $t$ yields the fundamental MeanFlow identity:
\begin{equation}
    u(z_t, r, t) = v(z_t, t) - (t-r)\frac{d}{dt}u(z_t, r, t),
\end{equation}
where $\frac{d}{dt}u$ represents the total derivative along the trajectory, computed via a Jacobian-vector product (JVP). Model optimization is performed by regressing the parameterized function $u_\theta$ onto the right-hand side of this identity, applying a stop-gradient operation ($\mathrm{sg}$) to the target:
\begin{equation}
    \label{eq:MF}
    \mathcal{L}_{\mathrm{MF}} = \big\lVert u_\theta(z_t, r, t) - \mathrm{sg}\left[\hat{v} - (t-r)\frac{d}{dt}u_\theta\right]\big\rVert^2_{\mathrm{a}},
\end{equation}
where $\lVert \cdot \rVert_{\mathrm{a}}$ denotes the adaptive $\ell_2$ loss introduced in~\cite{meanflow}, and $\hat{v}$ incorporates classifier-free guidance~\cite{cfg} into the velocity target. 
Following training, a final sample can be generated utilizing a sampling step: $\hat{x} = \epsilon - u_\theta(\epsilon, 0, 1)$.
More generally, generation can be executed in $n$ steps by integrating $u_\theta$ over a partition of the interval $[0, 1]$. 

MeanAudio \cite{meanaudio}, a model utilizing the MeanFlow objective \cite{meanflow}, achieves relatively better one-step audio generation quality compared to previous few-step sampling models~\cite{audiodear, audioturbo, consistencytta,audiolcm}, but still falls significantly behind its multi-step counterpart. In this work, we aim at reaching state-of-the-art performance for one-step T2A generation while closing the quality gap between one-step and multi-step sampling models.

\section{Method}
\label{sec:method}
\noindent Our goal is high-quality one-step T2A generation that keeps the inference efficiency of single-step sampling without giving up the option of multi-step sampling. We build on Fr\'echet-distance (FD) post-training, which Section~\ref{subsec:fd-post-training} formalizes together with its multi-representation form. As discussed there, FD post-training is an effective one-step objective but overwrites the model's learned velocity field, so the post-trained model can no longer sample well across multiple steps. To support both one- and multi-step sampling, Section~\ref{sec:anchor} adds a MeanFlow consistency term that anchors the FD distribution matching while preserving the multi-step trajectory.

\subsection{Fr\'echet-distance post-training}
\label{subsec:fd-post-training}
\noindent \textbf{Fr\'echet-distance (FD).} 
To evaluate how closely a generative model replicates real-world data, models are often compared at a distribution level within a fixed representation space. Let $\psi(\cdot)$ represent a pretrained, frozen representation extractor. We map a collection of reference target audio clips, $\mathcal{A}_{\text{gt}} = \{a_i\}$, and a corresponding set of generated audio samples, $\mathcal{A}_{\text{g}} = \{\hat{a}_i\}$, through this extractor to obtain their latent vectors. Assuming these embeddings follow multivariate normal distributions, we compute their empirical first and second moments:
\begin{equation}
    \mu_{\text{gt}} = \mathbb{E}[\psi(a)], \quad \Sigma_{\text{gt}} = \text{Cov}[\psi(a)]
\end{equation}
\begin{equation}
    \mu_{\text{g}} = \mathbb{E}[\psi(\hat{a})], \quad \Sigma_{\text{g}} = \text{Cov}[\psi(\hat{a})]
\end{equation}
The structural discrepancy between these two continuous domains is quantified using the Wasserstein-2 distance, expressed here as the Fr\'echet metric:
\begin{equation}
    \label{eq:fd-metric}
    FD_{\psi}(\mathcal{A}_{\text{gt}},\mathcal{A}_{\text{g}}) = \|\mu_{\text{gt}}-\mu_{\text{g}}\|_{2}^{2} + \text{Tr}\big(\Sigma_{\text{gt}}+\Sigma_{\text{g}}-2(\Sigma_{\text{gt}}\Sigma_{\text{g}})^{\frac{1}{2}}\big)
\end{equation}
When $\psi$ is instantiated as the VGGish architecture~\cite{vggish}, this metric resolves to the standard Fr\'echet Audio Distance (FAD)~\cite{fad}.

\input{figures/diagram}

\textbf{FD-loss.} 
Since FD quantifies the distribution-wise discrepancy between generated audio ($\mathcal{A}_{\text{g}}$) and ground-truth audio ($\mathcal{A}_{\text{gt}}$), it serves as a natural and direct training objective for generation tasks, denoted as $\mathcal{L}_{FD}$ in Eq.~\eqref{eq:fd}. 
\begin{equation}
    \label{eq:fd}
    \mathcal{L}_{FD} = FD_{\psi}(\mathcal{A}_{\text{gt}},\mathcal{A}_{\text{g}})
\end{equation}
However, treating this evaluation metric as an explicit optimization loss presents major computational problems. Computing a reliable, full-rank covariance matrix demands tracking thousands of generated instances across the training trajectory. Backpropagating gradients through such a large population pool at every training iteration is computationally prohibitive, while attempting to estimate these moments over small training mini-batches introduces extreme statistical variance that destabilizes the training process. To bypass these limitations for FD estimation, \cite{fdloss} proposed a queue-based estimator by using a queue that caches the representation vectors of the $N$ most recently synthesized audio clips. At each optimization step, the representations extracted from the active batch push out the oldest $B$ elements in the queue. The empirical mean and covariance are then evaluated over the entire active queue. Crucially, elements inherited from prior iterations are detached from the computational graph, enforcing that gradients flow only through the vectors belonging to the current batch. 


\textbf{Multi-representation FD-loss.} 
Relying exclusively on a single feature extractor may fail to capture the full spectrum of perceptual audio characteristics. The FD-loss framework can be naturally extended to optimize the generator across a diverse ensemble of $J$ independent representation spaces, denoted as $\{\psi_j\}_{j=1}^{J}$. A practical challenge in this joint optimization is that the raw Fr\'echet Distances computed in different embedding spaces often vary by several orders of magnitude. A naive summation would cause the representations with naturally larger numerical scales to disproportionately dominate the learning signal. To ensure stable and balanced optimization, we dynamically normalize each representation's contribution to a unit scale. This is achieved by dividing each loss term by its stop-gradiented current value, accompanied by a small stability constant $c$:
\begin{equation}
\label{eq:multi-fd}
    \mathcal{L}_{\text{Multi-FD}} = \sum_{j=1}^{J} w_j \frac{FD_{\psi_j}(\mathcal{A}_{\text{gt}}, \mathcal{A}_{\text{g}})}{\text{sg}\big(FD_{\psi_j}(\mathcal{A}_{\text{gt}}, \mathcal{A}_{\text{g}})\big) + c}
\end{equation}
where $w_j$ designates the specific weighting factor for the $j$-th representation space. This dynamic scaling standardizes the gradient magnitudes across all representation domains. 

While FD-loss successfully aligns the generated distribution with the target manifold, it inherently forces the model into a single-step generation paradigm. During post-training, the generated sample $\hat{x}$ must be synthesized via a single sampling step directly from noise, bypassing the iterative numerical solver. Attempting to optimize FD-loss on outputs generated via multiple sampling steps is impractical; backpropagating gradients through a fully unrolled $T$-step generation trajectory requires caching intermediate network activations across all steps, vastly exceeding standard hardware memory limits. Consequently, the optimization of FD post-training updates the network weights to map pure noise directly to the clean data distribution in a single forward pass, overwriting the model's originally learned progressive velocity field. This fundamentally reshapes the model from a continuous flow integrator into a rigid, one-step distribution mapper, resulting in severe audio degradation if the post-trained model is subsequently evaluated using a multi-step sampling process.

\subsection{MeanFlow anchor for multi-step preservation}
\label{sec:anchor}

\noindent To resolve the multi-step collapse induced by standard FD post-training, the optimization process must be constrained so that the model cannot simply discard its learned velocity field in favor of a rigid, one-step mapping. To achieve this, as illustrated in Fig.~\ref{fig-diagram} (b), we propose anchoring the FD distribution-matching process with the MeanFlow~\cite{meanflow} objective.

While the FD-loss evaluates the model at the terminal timestep to pull the one-step output towards the real data manifold, the MeanFlow objective $\mathcal{L}_{\mathrm{MF}}$ in Eq.~\eqref{eq:MF} explicitly enforces trajectory consistency across arbitrary sub-intervals $[r, t]$. By jointly optimizing these two objectives, the MeanFlow loss acts as a powerful regularizer. It forces the network to ensure that its predictions remain valid average velocities that can be numerically integrated, even as the FD-loss reshapes the global distribution of the terminal state. 

Formally, during the post-training phase, the network parameters $\theta$ are updated by minimizing a joint objective function that combines the single-step distributional alignment with the continuous trajectory constraint:
\begin{equation}
\label{eq:overall_obj}
    \mathcal{L}_{\text{total}} = \mathcal{L}_{\mathrm{\text{Multi-FD}}} + \lambda \mathcal{L}_{\mathrm{MF}}
\end{equation}
where $\lambda$ is a balancing hyperparameter. 

In practice, at each training iteration, the optimization is decoupled into two computational streams. First, the model calculates the MeanFlow loss $\mathcal{L}_{\mathrm{MF}}$ over uniformly sampled intermediate timesteps $r < t \sim \mathcal{U}(0,1)$ to maintain the integrity of the progressive denoising path. Second, the model evaluates a separate one-step forward pass directly from noise to data ($r=0, t=1$) to compute the generated sample $\hat{a}$ required for the FD-loss calculation. This dual-stream approach guarantees that the model simultaneously learns to produce one-step audio generation via FD-loss, while fundamentally preserving its identity as a continuous flow integrator capable of high-fidelity multi-step sampling.

\section{Experimental Setup}
\label{sec:setup}
\subsection{Data}
\noindent The training data we used is an aggregation of AudioCaps~\cite{audiocaps} and WavCaps~\cite{wavcaps}, resulting in a total of 387,438 audio samples. Due to budget restrictions, we only sampled 80,000 samples randomly and trained each model for one epoch.
Evaluation is conducted on the AudioCaps test set 10-second clips.

\subsection{Metrics} 
\noindent Several key objective and subjective metrics are used to evaluate the performance of the synthesized audio. To measure fidelity at the distribution level, we include two metrics: $\text{FD}_{\text{PANN}}$ and FAD~\cite{fad}. Both metrics calculate the Fréchet Distance via Eq.~\eqref{eq:fd-metric}, but utilize different backbone representation extractors. In this work, $\text{FD}_{\text{PANN}}$ leverages PANNs~\cite{panns} (denoted as FD), whereas FAD employs VGGish~\cite{vggish}.
We also include the Kullback–Leibler (KL) divergence and the Inception Score (IS;~\cite{is}) to evaluate the distribution of sound events and overall generation diversity. To measure how well a T2A model follows the semantic content of a natural language prompt, we report the CLAP score~\cite{clap}\footnote{CLAP model: \href{https://huggingface.co/lukewys/laion_clap/blob/main/music_speech_audioset_epoch_15_esc_89.98.pt}{music\_speech\_audioset\_epoch\_15\_esc\_89.98.pt}.}. To assess human perceptual quality, we conduct subjective listening tests to gather Mean Opinion Scores. Specifically, 9 raters evaluated a total of 90 generated audio samples. These include REL to evaluate text-relevance, indicating how faithfully the audio aligns with the prompt, and OVL to measure overall audio quality. 
To measure the inference speed of T2A models, we follow IMPACT~\cite{impact} by reporting the time to generate eight 10-second audio clips on an NVIDIA V100 32GB RAM GPU.

\input{tables/main_results}
\input{tables/subjective_eval}

\subsection{Training}
  \noindent Throughout all experiments, we initialize from the pretrained \textbf{MeanAudio-S-Full} checkpoint\footnote{MeanAudio-S-Full model checkpoint: \href{https://huggingface.co/AndreasXi/MeanAudio/blob/main/meanaudio_s_full.pth}{meanaudio\_s\_full.pth}}, a MeanFlow-based one-step T2A generator, and post-train its backbone with our joint objective in Eq.~\eqref{eq:overall_obj}. The autoencoder, BigVGAN vocoder, and the text encoders (FLAN-T5~\cite{flan-t5} and LAION-CLAP~\cite{clap}) are kept frozen throughout. Only the generator is updated.

  For the multi-representation FD-loss in Eq.~\eqref{eq:multi-fd} we use four complementary pretrained audio encoders: PANNs~\cite{panns}, PaSST~\cite{passt}, BEATs~\cite{beats}, and AudioMAE~\cite{audiomae}, aggregated with
  uniform weights $w_j{=}1$ and the stability constant $c{=}0.01$. Notably, the VGGish encoder is excluded from this ensemble to ensure that we are not directly optimizing the FAD metric.
  We adopt the queue-based estimator in Section~\ref{subsec:fd-post-training} with a queue size of $N{=}20{,}000$; the reference moments $\mu_{\mathrm{gt}},\Sigma_{\mathrm{gt}}$ are precomputed once per encoder over the training audio. At each step, the FD-loss is evaluated on a one-step generated sample synthesized directly from noise ($r{=}0,t{=}1$), while the MeanFlow anchor $\mathcal{L}_{\mathrm{MF}}$ is
  computed on an independently drawn pair $r{<}t\sim\mathcal{U}(0,1)$; the anchor weight is set to $\lambda{=}0.25$.

  Optimization uses AdamW with weight decay $10^{-6}$ and a constant learning rate of $1{\times}10^{-5}$, 200 linear warm-up steps, and gradient-norm clipping at $1.0$, using a batch size of $2$. We maintain an exponential moving average (EMA) of the model weights and use the EMA weights for all evaluations, selecting the checkpoint with the best validation FD score. Classifier-free guidance is folded into the MeanFlow velocity target during training, so one-step inference is a single conditional forward pass with no separate unconditional evaluation; $n$-step generation integrates $u_\theta$ over a uniform partition of $[0,1]$.

\section{Results}
\label{sec:results}

\subsection{System-level comparisons}
\noindent To position our work within the broader landscape of text-to-audio generation, we compare the system-level performance of FdAudio against various state-of-the-art multi-step and few-step models on the AudioCaps evaluation set.

Among few-step methods, FdAudio establishes state-of-the-art one-step generation performance. In a strictly one-step sampling scenario, FdAudio achieves an FD of 12.71, an FAD of 1.26, and an IS score of 11.14. This significantly outperforms its direct baseline, MeanAudio-S-Full, which yields an FD of 14.35 and an FAD of 1.77 for one-step generation. Furthermore, FdAudio surpasses other one-step models, such as AudioLCM, ConsistencyTTA, SoundCTM, and AudioDEAR across primary objective metrics.

When evaluated using a 25-step sampling trajectory, FdAudio maintains highly competitive generation quality. Its multi-step performance is comparable to or surpasses several strong multi-step systems, including TangoFlux, EzAudio-XL, and IMPACT, despite FdAudio utilizing a relatively compact parameter count of 120M. Under this 25-step setting, FdAudio achieves an improved FD of 12.56, an IS of 11.68, and a KL divergence of 1.23.

Beyond objective metrics, we further validate our approach through subjective human evaluation, focusing on text-relevance (REL) and overall audio quality (OVL). As shown in Table~\ref{tab:rel_ovl_scores}, FdAudio consistently outperforms MeanAudio in human perception across both sampling budgets. 
Under the strict one-step setting, FdAudio improves the REL score from $4.10$ to $4.24$ and the OVL score from $3.81$ to $3.93$ compared to MeanAudio. This trend also holds in the 25-step regime. These perceptual gains support our objective findings, demonstrating that the MeanFlow-anchored FD post-training effectively enhances both one-step and multi-step generation quality.

In terms of inference efficiency, generating a batch of eight 10-second clips requires only 1.79 seconds for the one-step FdAudio model. When utilizing the 25-step sampling process, the latency remains highly efficient at 3.21 seconds. In stark contrast, prominent multi-step models such as Tango 2 and AudioLDM2-L demand 182.23 seconds and 194.77 seconds, respectively, to process an identical batch size.

\input{tables/MF-anchor-weight}

\subsection{MeanFlow anchor}
\label{sec:mf-anchor}
\noindent Crucially, our proposed MeanFlow anchor prevents the multi-step collapse typically caused by naive FD post-training. The original FD post-training forces the model into a rigid, single-step distribution mapper, overwriting its originally learned progressive velocity field. By anchoring the optimization with the MeanFlow objective, we explicitly enforce trajectory consistency, preserving the network's ability to act as a continuous generator capable of high-fidelity multi-step sampling.

Table~\ref{tab:mf-anchor-weight} quantitatively demonstrates the critical importance of this training objective. Without the MeanFlow anchor ($\lambda=0.00$), the one-step performance remains reasonable, achieving an FAD of 1.27. However, attempting to evaluate this unanchored model using a 25-step sampling trajectory results in severe degradation on fidelity, with the FAD score degrading sharply to 3.61 and the FD score worsening to 15.13. With the MeanFlow anchor ($\lambda=0.25$), the model successfully maintains its multi-step generation capabilities without severely sacrificing one-step fidelity. In fact, the anchored model ($\lambda=0.25$) slightly improves one-step generation, yielding an IS of 11.14 compared to 10.51 for the unanchored baseline ($\lambda=0.00$), while achieving a highly competitive 25-step FD of 12.56 and an IS of 11.68. 
This confirms that our method makes one-step generation highly competitive while retaining the multi-step path as an option for applications that prioritize maximum fidelity and diversity.

Furthermore, we evaluate the impact of the MeanFlow anchor's balancing weight ($\lambda$) on generation quality, as detailed in Table~\ref{tab:mf-anchor-weight}. A weight of $0.25$ provides the optimal balance, yielding the strongest overall performance across both the one-step and 25-step generation bounds. Increasing the weight to 0.50 slightly improves the one-step FAD to 1.15, but it introduces noticeable degradations in the IS score, which drops to 10.22 for one-step and 11.17 for 25-step generation. Finally, applying a heavy MeanFlow penalty at a weight of 1.00 over-regularizes the network, worsening both one-step and multi-step FAD metrics to 1.51 and 1.66, respectively, while further depressing the IS score.

\input{tables/encoder-ablation}

\subsection{Encoders for multi-representation FD-loss}
\noindent To capture a broader spectrum of perceptual audio characteristics, we investigate the effect of different combinations of representations from different encoders within the post-training framework. Table~\ref{tab:fd-encoders} details the ablation studies exploring these pretrained audio encoder combinations utilized in the multi-representation FD-loss in Eq.~\eqref{eq:multi-fd}. 

When evaluating the system under a strict one-step sampling budget, the combination of PANNs and PaSST serves as our baseline, achieving an FD score of 13.11 and an FAD of 1.25. Incorporating the BEATs encoder yields a lower FD score of 12.91 and improves the FAD score to 1.17, though the IS score drops slightly to 10.37. Conversely, combining PANNs and PaSST with AudioMAE noticeably benefits the CLAP metric, raising it to 0.356, but causes the FAD score to degrade to 1.53. By integrating all four encoders, PANNs, PaSST, BEATs, and AudioMAE, the model achieves a highly balanced one-step performance profile, yielding the lowest FD score of 12.71, an FAD score of 1.26, and an IS score of 11.14, while maintaining a strong CLAP score of 0.345.

A similar trend is observed when evaluating the model across a 25-step sampling trajectory. The full ensemble of PANNs, PaSST, BEATs, and AudioMAE provides the lowest overall FD score at 12.56 and a competitive FAD score of 1.59. While specific sub-combinations can favor individual metrics, such as the PANNs + PaSST + BEATs configuration reaching a lower 25-step FAD score of 1.38, the four-encoder configuration consistently ensures robust generalization. This demonstrates that optimizing across a multi-representation ensemble effectively prevents the generator from overfitting to a single feature extractor's embedding space.

\section{Conclusion}
\label{sec:conclusion}
\noindent In this work, we introduced FdAudio, a novel post-training framework that bridges the quality gap between one-step and multi-step text-to-audio systems without sacrificing sampling flexibility. While direct Fréchet-distance (FD) post-training enhances single-pass distribution mapping, it destroys the model's progressive velocity field, causing severe multi-step collapse during iterative sampling. We successfully resolved this limitation by incorporating the MeanFlow consistency objective as an anchor.
This unified approach effectively closes the quality gap between one-step and multi-step models, while preserving the flexibility of the multi-step sampling trajectory for applications that require maximum acoustic fidelity and generation diversity.
Ablation studies confirm that optimizing across a multi-representation ensemble of four encoders prevents feature overfitting. Benchmarking on the AudioCaps dataset demonstrates that FdAudio establishes state-of-the-art one-step generation quality among few-step systems, yielding an 11.4\% reduction in FD score and a 28.8\% improvement in FAD score relative to the baseline MeanAudio framework. 

\bibliographystyle{IEEEtran}
\bibliography{ref}

\end{document}

%% file: figures/diagram.tex
\begin{figure*}[t]
    \centering
    \includegraphics[width=18.1cm]{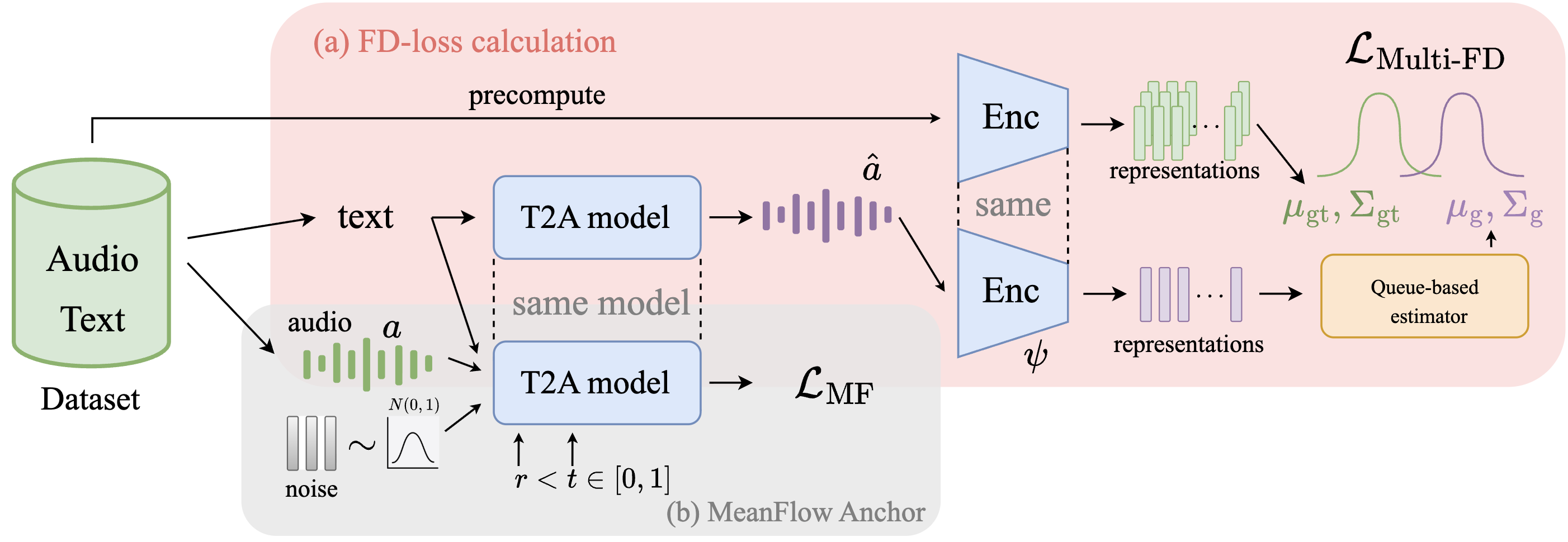}
    \caption{Overview of the post-training framework for FdAudio. (a) FD-Loss calculation pipeline: Ground truth audio statistics ($\mu_{gt}$, $\Sigma_{gt}$) are precomputed prior to training. During the optimization step, a one-step forward pass synthesizes a generated sample $\hat{a}$. This sample is processed through frozen, pretrained encoders ($\psi$) to compute and maintain the generated statistics ($\mu_{g}$, $\Sigma_{g}$) via a queue-based estimator. (b) MeanFlow Anchor: To prevent the multi-step collapse associated with naive FD post-training, the framework computes the MeanFlow consistency loss ($\mathcal{L}_{MF}$) over uniformly sampled intermediate timesteps ($r < t \in [0, 1]$), preserving the continuous sampling trajectory for high-fidelity multi-step generation. }
    \label{fig-diagram}
\end{figure*}

%% file: tables/main_results.tex
\begin{table*}[ht]
    \centering
    \caption{System-level performance of T2A generation models on AudioCaps.
    ``Step'' is the number of sampling steps; 
    Latency is the time to generate a batch of 8 clips on an Nvidia V100 32GB.
    Best values among few-step methods are \textbf{bold}, second-best \underline{underlined}.}
    \label{tab:sota}
    \setlength{\tabcolsep}{13pt}
    \renewcommand{\arraystretch}{0.9}
    \begin{tabular}{lcc|ccccc|c}
    \toprule
    \textbf{AudioCaps} & \textbf{\# para} & \textbf{Step} & FD$\downarrow$ & FAD $\downarrow$ & KL$\downarrow$ & IS$\uparrow$ & CLAP$\uparrow$ & Latency$\downarrow$ \\
    \midrule
    Ground Truth & - & - & - & - & - & - & 0.412 & - \\
    \midrule
    \multicolumn{9}{l}{\textbf{Multi-step Sampling}}\\
    Tango 2        & 866M & 200 & 17.60 & 2.82 & 1.18 & 10.12 & 0.328  & 182.23 \\
    TangoFlux      & 516M & 50  & 19.69 & 2.31 & 1.19 & 12.35 & 0.318  & 45.50 \\
    EzAudio-XL (24kHz)     & 874M & 50 & 13.94 & 3.30 & 1.27 & 11.37 & 0.314 & 39.78 \\
    Make-an-Audio 2& 160M & 100 & 14.76 & 1.21 & 1.27 & 9.99 & 0.282   & 15.87 \\
    AudioLDM2-L    & 712M & 200 & 24.20 & 2.10 & 1.54 & 8.31 & 0.232   & 194.77 \\
    AudioMNTP      & 193M & 100 & 14.81 & 1.68 & 1.16 & 9.67 & -       & - \\
    IMPACT         & 193M & 100 & 15.25 & 1.26 & 1.06 & 10.57 & -      & 22.34 \\
    Resonate-GRPO (44kHz) & 468M & 25 & 15.04 & 2.26 & 1.30 &  10.60 & 0.420 & 18.80 \\
    \midrule
    MeanAudio-S-Full & 120M & 25 & 13.42 & 2.31 & 1.26 & 11.23 & 0.323 & 3.21 \\
    \textbf{FdAudio (ours)} & 120M & 25 & 12.56 & 1.59 & 1.23 & 11.68 & 0.344 & 3.21 \\
    \midrule
    \midrule
    \multicolumn{9}{l}{\textbf{Few-step Sampling}}\\
    ConsistencyTTA & 559M & 1 & 21.74 & 2.48 & 1.39 & 8.91 & 0.270 & 3.03 \\
    SoundCTM       & 868M & 1 & 21.02 & 2.43 & 1.42 & 8.00 & 0.246 & 2.48 \\
    AudioLCM       & 160M & 1 & 26.46 & 4.21 & 1.62 & 6.60 & 0.190 & 2.75 \\
    AudioLCM       & 160M & 2 & 21.26 & 2.02 & 1.40 & 8.26 & 0.225 & 2.93 \\
    AudioTurbo     & 1.1B & 5  & 22.18 & - & \underline{1.30} & 8.88 & - & - \\
    AudioDEAR      & 191M & 1 & 18.67 & 2.79 & \textbf{1.06} & 9.66 & - & 2.61 \\
    \midrule
    MeanAudio-S-Full & 120M & 1 & \underline{14.35} & \underline{1.77} & 1.32 & \underline{10.43} & \underline{0.317} & \textbf{1.79} \\
    \textbf{FdAudio (ours)} & 120M & 1 & \textbf{12.71} & \textbf{1.26} & \underline{1.30} & \textbf{11.14} & \textbf{0.345} & \textbf{1.79} \\
    \bottomrule
    \end{tabular}
    \vspace{-1em}
\end{table*}

%% file: tables/subjective_eval.tex
\begin{table}[ht]
    \centering
    \caption{Mean and standard error values of subjective evaluation for MeanAudio and FdAudio at one-step and 25-step.}
    \label{tab:rel_ovl_scores}
    \setlength{\tabcolsep}{10pt}
    \renewcommand{\arraystretch}{0.9}
    \begin{tabular}{lccc}
    \toprule
    \textbf{Model} & \textbf{Step} & \textbf{REL$\uparrow$} & \textbf{OVL$\uparrow$} \\
    \midrule
    Ground Truth & - & $4.40 \pm 0.08$ & $4.23 \pm 0.09$ \\
    \midrule
    MeanAudio-S-Full & 25 & $4.21 \pm 0.08$ & $4.00 \pm 0.07$ \\
    \textbf{FdAudio (ours)} & 25 & $\textbf{4.33} \pm 0.10$ & $\textbf{4.13} \pm 0.09$ \\
    \midrule
    MeanAudio-S-Full & 1 & $4.10 \pm 0.09$ & $3.81 \pm 0.11$ \\
    \textbf{FdAudio (ours)} & 1 & $\textbf{4.24} \pm 0.10$ & $\textbf{3.93} \pm 0.09$ \\
    \bottomrule
    \end{tabular}
    \vspace{-1em}
\end{table}

%% file: tables/MF-anchor-weight.tex
\begin{table}[th]
    \centering
    \caption{Ablation studies on the weights of the MeanFlow anchor. 
    }
    \label{tab:mf-anchor-weight}
    \small
    \renewcommand{\arraystretch}{0.9}
    \begin{tabular}{cc|ccccccc}
    \toprule
    $\mathbf{\lambda}$ & \textbf{Step} & FD$\downarrow$ & FAD$\downarrow$ & KL$\downarrow$ & IS$\uparrow$ & CLAP$\uparrow$ \\

    \midrule
    0.00 & 1 & 13.04 & 1.27 & 1.32 & 10.51 & 0.344 \\
    0.25 & 1 & \textbf{12.71} & 1.26 & 1.30 & \textbf{11.14} & \textbf{0.345} \\
    0.50 & 1 & 13.84 & \textbf{1.15} & \textbf{1.28} & 10.22 & 0.334 \\
    1.00 & 1 & 14.59 & 1.51 & 1.30 & 10.26 & 0.321 \\
    
    \midrule
    0.00 & 25 & 15.13 & 3.61 & 1.29 & 11.52 & 0.342 \\
        0.25 & 25 & \textbf{12.56} & 1.59 & \textbf{1.23} & \textbf{11.68} & \textbf{0.344} \\
    0.50 & 25 & 14.16 & \textbf{1.55} & 1.29 & 11.17 & 0.335 \\
    1.00 & 25 & 14.39 & 1.66 & 1.27 & 11.06 & 0.333 \\
    \bottomrule
    \end{tabular}
    \vspace{-1.5em}
\end{table}

%% file: tables/encoder-ablation.tex
\begin{table}[h]
      \centering
      \caption{Ablation studies of different encoder combinations for the multi-representation FD-loss.}
      \label{tab:fd-encoders}
      \scriptsize
      \begin{tabular}{lc|ccccccc}
      \toprule
      \textbf{Encoders} & \textbf{Step} & FD$\downarrow$ & FAD$\downarrow$ & KL$\downarrow$ & IS$\uparrow$ & CLAP$\uparrow$ \\
      \midrule
      PANNs+PaSST            & 1 & 13.11 & 1.25 & 1.30 & 10.78 & 0.326 \\
      \quad+BEATs           & 1 & 12.91 & \textbf{1.17} & 1.29 & 10.37 & 0.335 \\
      \quad+AudioMAE        & 1 & 13.75 & 1.53 & \textbf{1.26} & 11.06 & \textbf{0.356} \\
      \quad+BEATs+AudioMAE  & 1 & \textbf{12.71} & 1.26 & 1.30 & \textbf{11.14} & 0.345 \\
      \midrule
      PANNs+PaSST            & 25 & 12.74 & 1.48 & 1.24 & 11.67 & 0.336 \\
      \quad+BEATs           & 25 & 12.87 & \textbf{1.38} & \textbf{1.22} & 11.56 & 0.339 \\
      \quad+AudioMAE        & 25 & 12.81 & 1.63 & 1.24 & \textbf{11.70} & 0.341 \\
      \quad+BEATs+AudioMAE  & 25 & \textbf{12.56} & 1.59 & 1.23 & 11.68 & \textbf{0.344} \\
      \bottomrule
      \end{tabular}
      \vspace{-1em}
  \end{table}